\documentclass[aps,prc,showpacs,float,preprint]{revtex4-1}
\usepackage{graphicx}
\usepackage[section]{placeins}
\begin{document}
\title{Analytical Solutions of the Quantum Hamilton-Jacobi Equation and Exact WKB-Like Representations of  One-Dimensional Wave Functions}
\author{Mario Fusco Girard}
\affiliation{Department of Physics ''E.R. Caianiello'',\\University of Salerno \\and \\Gruppo Collegato INFN di Salerno,\\Via Giovanni Paolo II, 84084 Fisciano (SA), Italy\\
electronic address: mario@sa.infn.it}

\begin{abstract}
General analytical solutions of the Quantum Hamilton Jacobi Equation for conservative one-dimensional or reducible motion are presented and discussed. The quantum Hamilton's characteristic function and its derivative, i.e. the quantum momentum function, are obtained in general, and it is shown that any one-dimensional wave function can be exactly represented in a WKB-like form. The formalism is applied to the harmonic oscillator and to the electron's motion in the hydrogen atom, and the above mentioned functions are computed and discussed for different quantum numbers. It is analyzed how the quantum quantities investigated tend to the corresponding classical ones, in the limit $\hbar\to 0$. These results demonstrate that the Quantum Hamilton Jacobi Equation is not only completely equivalent to the Schr\"odinger Equation, but allows also to fully investigate the transition from quantum to classical mechanics.
\end{abstract}
\maketitle

\section{Introduction}
The Hamilton's characteristic function [1] also called reduced action [2], i.e. the solution of the time independent Hamilton-Jacobi equation, has a very important role in the classical mechanics of conservative systems. Indeed, it allows either to completely solve the dynamical problem, in the case of a complete integral, or to investigate the properties of iso-energetic families of trajectories, in the case of particular integrals [2-7]. It is therefore of interest to investigate what is the quantum quantity that in the classical limit generates this function. This study cannot be done by means of the Schr\"odinger Equation (SE), which for $\hbar\to 0$ loses its significance [8]. In order to explore the classical limit, it is necessary to appeal to the Quantum Hamilton-Jacobi Equation (QHJE), i.e. the equation for the time-independent part of the phase of the wave function. From the very beginning of the Quantum Mechanics, this equation was recognized as the natural bridge between classical and quantum world [9, 10]: indeed, in principle it is completely equivalent to the SE, and moreover, for $\hbar = 0$, it reduces to the classical Hamilton-Jacobi equation.
The search for solutions of the QHJE as a power series of $\hbar$ produced the WKB method [10, 11] in which the wave function is approximately expressed, at first orders of the expansion, by means of the classical reduced action and its gradient, i.e. the classical momentum. In the fifties, the QHJE was the starting point of the Bohm's interpretation of Quantum Mechanics [12, 13]. In more recent years, the interest for the QHJE was renewed by Leacock and Padgett [14, 15] who showed how it is possible to find the energy levels for one dimensional motion without explicitly solving the equation itself, and by Floyd [16-22] in the framework of a hidden variables theory. Bhalla, Sree Rajani, Kapoor and coworkers [23-28] extended the approach of Leacock and Padgett, showing that also the eigenfunctions for a wide class of potentials can be found by means of the QHJE. Further results in this direction were found by Yesiltas [29] and Bouda [30, 31]. The QHJE equation was also derived by Faraggi and Matone [32-34] from of an equivalence principle, within a geometric-differential approach. A QHJE was obtained by Perival [35], for the generating function of a canonical transformation of phase space integral. An operatorial QHJE was investigated by Roncadelli and Shulmann [36]. Some geometrical aspects of the connection between Hamilton-Jacobi theory and Quantum Mechanics were studied by Marmo et al. [37].
Returning to the WKB approximation, it is well known that the series does not converge and moreover, the expansion fails near the classical turning points. But above all, not even in this approach it is possible to identify the quantum quantities which in the limit $\hbar \to 0$ generate the classical action and momentum. This requires the knowledge of the exact solutions of the QHJE. Particular solutions of that equation can be found as demonstrated by Bhalla and coworkers, and are the same obtained from the complex logarithm of the solution of the Schr\"odinger equation at that energy. However, in the classically allowed regions, it is not possible to put directly in relation these quantities with the corresponding classical ones. Indeed, those solutions appear to be imaginary for each value of the coordinate, and have discontinuities in correspondence of the nodes of the wave function; these singularities do not vanish for $\hbar\to 0$, and their number goes to infinite in this limit. In the classically allowed regions, the classical quantities are instead real and continuous. In these regions therefore we need to construct more general, complex and continuous solutions of the QHJE, whose imaginary part vanishes in the classical limit. Nevertheless these more general solutions must correspond to the same wave functions, as the particular solutions; moreover they must reduce to these latter in the classically forbidden regions, where also the classical quantities are purely imaginary. Actually, as it will be shown in the following, to the same wave function it corresponds a whole family of complex solutions of the QHJE, which reduce to the imaginary one for a particular choice of the parameters. In a previous paper [38] these complex solutions were found by making use of numerical integration of the QHJE. In the present paper it is shown how to analytically find them, by means of general integrals of the QHJE, obtained by exploiting known results of the theory of the Riccati differential equation. In the following it will be shown that the real parts of these complex solutions in the limit $\hbar \to 0$ generate the classical characteristic function and the corresponding momentum in the classically allowed regions, while the imaginary parts correctly vanish there. Moreover, it is found that every one-dimensional wave function can be expressed in an exact WKB-like form, while this representation is not possible by means of the imaginary particular solutions. Finally, in this approach the exact generalization of the semiclassical quantization rule is obtained. The results presented here demonstrate that the approach to Quantum Mechanics based on the QHJE not only is completely equivalent to one based on the Schr\"odinger Equation, but also it allows to fully clarify the transition from quantum and classical mechanics.

This paper is so organized: in Sect. 1 the method to construct the suitable general solutions of the QHJE for the one-dimensional case is exposed. Sect. 2 is devoted to the presentation of the results and some comments for the case of the harmonic oscillator. Sect. 3 refers to the case of the electron in the hydrogen atom. Sect. 4 is devoted to a discussion of the semiclassical limit. Some conclusions are reported in Sect. 5.

\section{THE METHOD}
Our starting point, as in the usual WKB method, is the search for solutions of the form:
\begin{equation}
\psi(x,E)=Ae^{{i\over \hbar} W(x,E)}
\end{equation}	
of the time-independent SE for one dimensional motion of a particle of mass m and energy E in a potential $V(x)$:
 \begin{equation}
-{\hbar ^2\over 2m} {d^2\psi\over dx^2}=\left[E-V(x)\right]\psi\ .
\end{equation}    
Substitution of Eq. (1) in (2) gives the one-dimensional time-independent QHJE:
\begin{equation}
{1\over 2m}\left({dW\over dx}\right)^2 - {i\hbar \over 2m}{d^2W\over dx^2}=E - V(x)\ .
\end{equation}  
The function $W(x, E)$, which is proportional to the complex logarithm of the wave function, is the quantum characteristic function (or quantum reduced action) of the particle.
For $\hbar = 0$,  Eq. (3) is the classical Hamilton-Jacobi equation  [1] for the Hamilton's characteristic function $W_C(x,E)$:
\begin{equation}
{1\over 2m}\left({dW_C\over dx}\right)^2 = E - V(x)\ ,
\end{equation}  
whose solutions are
\begin{equation}
W_C(x,E) = \int p_c(x,E) dx = \pm\int \sqrt{2m \left({E- V(x)}\right)} dx \ ;
\end{equation} 
here $p_C(x,E)$ is the particle's classical momentum.
From Eq. (1), it follows that
\begin{equation}
 p_F(x,E) = W'(x,E) = {\hbar\over i} {\psi '(x,E)\over \psi(x,E)}
\end{equation} 
(apex  denoting derivative with respect to the argument x). $p_F (x,E)$ is called the quantum momentum function [12].  For a real wave function, $p_F(x,E)$ as given from Eq. (6) is purely imaginary, with first order poles in correspondence of the nodal points.
Eq. (3) can also be written as a first order Riccati equation for $p_F(x, E)$:
\begin{equation}
 p_F' = {1\over i\hbar} (p_F)^2-{2m\over i\hbar}(E-V(x))\ .
\end{equation}    
Given a solution $p_F(x, E)$ of this equation, the corresponding quantum characteristic function is obtained by integration:
\begin{equation}
W(x,E) = \int p_F(x,E) dx\ .
\end{equation} 
 Skipping the trivial case of a free particle, in this paper we are interested to the one dimensional bound states in the potential $V(x)$. The wave functions are then real functions, oscillating in the classically allowed regions $(E\ge V(x))$ and vanishing for $|x|\to \infty$.
Leacock and Padgett [12, 13] demonstrated that the exact quantum energy levels can be obtained from the condition:
\begin{equation}
\oint p_F(x,E) dx = 2n\pi\hbar\ ,
\end{equation} 
where the integration is done along a closed path in the complex x-plane, enclosing  the segment between the turning points. The integral can be connected to the energy as shown in [12, 13] for  various systems,  obtaining so the exact energy levels without integrating the QHJE itself.

Given an energy level satisfying eq. (9), a corresponding particular solution $p_{F,S}(x,E)$  of the Eq. (7) can be obtained as demonstrated in [25], by investigating the equation in the complex x-plane.

From $p_{F,S}(x,E)$, by means of Eq. (8) a particular solution $W_S (x,E)$ of the QHJE (3) is obtained. Both these functions can also be achieved by the complex logarithm of the wave function $\psi(x,E)$, and then by using Eq. (6). If we try to derive from these quantum quantities the corresponding classical ones, we find that in the classically forbidden regions (c.f.r.), the classical momentum $p_C(x,E)$ is imaginary and it is correctly obtained as the limit of $p_{F,S}(x,E)$ for $\hbar\to 0$, but this correspondence fails in the classically allowed regions (c.a.r.). Indeed, $p_{F,S}(x,E)$  as given from Eq. (6), is an imaginary quantity for each value of $x$, with poles at the nodal points of $\psi (x,E)$ (more exactly, it is a complex quantity with a real part everywhere zero, apart for delta singularities in correspondence of the nodal points). As for the corresponding quantum reduced action, $W_S(x,E)$ as computed from Eq. (1) is a complex quantity, with an imaginary part logarithmically diverging at the nodal points of $\psi(x,E)$, and a real part jumping from 0 to $\pi\hbar$ (mod $2\pi\hbar$) there, according to the sign of the wave function (these jumps give the delta singularities of the real part of $p_{F,S}(x,E)$). The singularities of $W_S(x,E)$, as well as those of $p_{F,S}(x,E)$, do not vanish in the limit $\hbar\to 0$, and their number moreover tends to infinite. The classical functions $W_C(x,E)$ and $p_C(x,E)$ are instead real continuous functions in the c.a.r. Therefore, $W_C(x,E)$ and $p_C(x,E)$ in the c.a.r. cannot be the limit of $W_S(x,E)$ and $p_{F,S}(x,E)$  for $\hbar\to 0$, but have instead to be the limit of the real parts of two continuous complex function $W_G(x,E)$ and $p_{F,G}(x,E)$, whose imaginary parts vanish when $\hbar\to 0$. As shown in the following, these functions can indeed be obtained from the general solutions of their equations (3) and (7) by means of known results for the Riccati equation; moreover, to these more general solutions corresponds the same wave function $\psi(x,E)$ which is connected to $W_S(x,E)$ and $p_{F,S}(x,E)$.

As we are looking for imaginary solutions in the c.f.r. and for complex ones in the c.a.r., the starting point is that the integration of Eqs. (3) and (7) has to be done separately in the various regions, and afterwards the solutions have to be matched together by imposing suitable continuity conditions. This is the same approach adopted in the usual WKB method. The crucial difference is that we will use exact solutions of the QHJE, instead of approximated ones.

Let us for simplicity suppose that the potential has only two turning points, $x_1$ and $x_2$, at the energy E, with $E\ge V(x)$  for $x_1\le x\le x_2$. Three regions are so defined: I):  $x<x_1$;  II):  $x_1\le x\le x_2$ ;   III): $x_2<x$. Regions I and III are the c.f.r., while region II is the c.a.r.

Given an allowed energy level from the condition (9), in the regions I and III we search a solution of Eq. (3) of the form:
\begin{equation}
W_S(x,E) = i Y_S (x,E)\ .
\end{equation}   
This is obtained by integrating the correspondent purely imaginary, particular solution $p_{F,S}(x,E)$ of Eq. (7),  found as demonstrated in [25], by Sree Rajani et al., where it is also shown that for most exactly solvable potentials, these particular solutions are rational functions.  For the low-lying states they can often be found by simple inspection.
Therefore, the wave functions in the c.f.r. I and III have the exact exponential-type representation:
\begin{equation}
\psi_{I,III}(x,E)=A_{I,III}e^{-Y_S(x,E)/\hbar}= A_{I,III}e^{-{1\over\hbar }\int p_{F,S}(x,E)dx}
\end{equation}   
here $A_{I,III}$ are suitable constants, to be fixed later.

As for the c.a.r. II, the purely imaginary particular solutions $p_{F,S}(x)$ and $W_S(x)=iY_S (x)$, already found for regions I and III, solve the respective Eqs. (7) and (3) also there.  However, according to the above considerations, we need for region II more general solutions, complex and continuous, corresponding to the same wave function $\psi(x, E)$, and from which it is possible to get the classical quantities in the limit $\hbar\to 0$.

(Hereafter the dependence of the various functions from the energy E will be often understood.)
In the c.a.r. we therefore write the phase $W(x)$ in the full complex form
\begin{equation}
W_G(x) =X(x) + i Y(x)\ .
\end{equation}             

By inserting this into Eq. (3) and by separately equating the real and imaginary parts of the two sides, we get
\begin{equation}
X'^2(x) - Y'^2(x) + \hbar Y''(x)  = 2m\left(E -V(x) \right)
\end{equation}   
\begin{equation}
X'(x) Y'(x) -{1\over 2} \hbar X''(x)  = 0 \ .
\end{equation}   
Equation (14) is immediately integrated:
\begin{equation}
Y(x)=\hbar \log\left[\sqrt | X'(x)|\right] \ .
\end{equation}   
As we will see, the integration constant here can be put equal to zero. This equation already suggests that the fundamental quantity is the real part $X(x)$ of the quantum characteristic function $W_G(x)$. By inserting last equation into Eq. (13), we get a nonlinear third order equation for $X(x)$:
\begin{equation}
 {4X'^4(x)-3\hbar ^2 X''^2(x) + 2\hbar^2X'(x)X'''(x)\over 4X'^2(x)} =2m(E-V(x))\ .
\end{equation}       
While in [38] Eq. (16) was numerically integrated, in the present paper we will recover the same results by computing $W_G(x)$ through Eq. (8), from the general integral of the Eq. (7). Analytical solutions of Eq. (16), depending on a couple of independent solutions of the SE were presented in [21, 22]. In the current paper, in order to demonstrate the equivalence of QHJE and SE, we will use a different approach, and we will construct analytical solutions of Eq. (16), without reference to solutions of the SE.

We will exploit a theorem on the Riccati differential equation [39], which states that if a particular solution $p_{F,S}(x)$ is known, the equation can be completely integrated and the general solution is given by:
\begin{equation}
p_{F,G}(x) = p_{F,S}(x) + {1\over v(x)}
\end{equation} 
where $v(x)$  is the general integral of an associated linear differential equation, containing $p_{F,S}(x)$. For Eq. (7), the associated equation is
\begin{equation}
v'(x) - \left({2i\over \hbar}\right)	p_{F,S}(x) v(x) ={i\over \hbar}
\end{equation}       
whose general solution is:
\begin{equation}
v(x) = { {i\over \hbar}\int _0^x e ^{-{2 i\over \hbar } \int_0^x p_{F,S}(x) dx} dx +C_0  \over e^{-{2i\over\hbar} \int_0^x p_{F,S}(x) dx} }
\end{equation} 

Here $C_0$ is a complex integration constant.

Thus, the general integral of the Riccati equation (7) is:
\begin{equation}
p_{F,G}(x) = p_{F,S}(x) + {e^{-{2i\over\hbar} \int_0^x p_{F,S}(x) dx} \over {i\over \hbar}\int _0^x e ^{-{2 i\over \hbar } \int_0^x p_{F,S}(x) dx} dx +C_0}
\end{equation} 
The second term in the r.h.s. of this equation will be denoted as $p_{F,add} (x)$.  Eq. (20) gives a family of complex solutions of the Eq. (7) depending on the parameter $C_0$, which in turn depends on the energy.  When $|C_0|\to\infty$, $p_{F,G} (x)$ reduces to the particular solution $p_{F,S}(x)$. In the following it will be discussed how to choice $C_0$ in order to get suitable solutions of Eq. (7), different from $p_{F,S}(x)$. While $p_{F,S}(x)$ is purely imaginary, the complex constant $C_0$ gives a real part to $p_{F,G} (x)$.
By integrating last equation and remembering Eq. (8) we get the general expression of the quantum characteristic function $W_G(x)$ in the classical region:
\begin{equation}
W_{G}(x) = W_{S}(x) +{\hbar\over i} \log \left[{i\over \hbar}\int_0^x  e^{-{2i\over\hbar} \int_0^x p_{F,S}(x) dx} dx + C_0\right] + C_1
\end{equation}   
here $C_1$ is another complex constant, depending on the energy.
The real part of this expression is:
\begin{equation}
X(x) = Re[W_S(x)] + \hbar Arg \left[{i\over \hbar}\int_0^x  e^{-{2i\over\hbar} \int_0^x p_{F,S}(x) dx} dx + C_0\right] + Re [C_1]
\end{equation}   
Here $Re$ denotes the real part, and $Arg$ the argument of the complex quantity inside the brackets.

In the following we will sometimes denote the second term in the r.h.s. of equations (21) and (22) as $W_{add}(x)$ and $X_{add}(x)$, respectively.
Eq. (22) is the general solution of the third order equation (16), which is equivalent to the Schr\"odinger equation. $C_0$ and $Re[C_1]$ give the three real parameters corresponding to the Cauchy data needed to obtain a particular solution of this equation.

From a solution $X(x)$  of Eq. (22), by using Eqs. (15), (12) and (1), we obtain a complex solution of the Schr\"odinger equation in the c.a.r., of the form
\begin{equation}
\psi_{M}=\frac{A_{II}}{\sqrt{|X'(x)|}}\exp\left[{iX(x)\over \hbar}\right]\ ;
\end{equation} 		     
here $A_ {II}$ is a complex constant, that in analogy with the WKB approach, we choose in the form
\begin{equation}
A_{II} = A e^{i\pi\over 4}\ ,
\end{equation} 				
where A is a real constant. By taking the imaginary part of $\psi_M(x)$, the wave function in the region II can be put in WKB-like form [10]:
\begin{equation}
\psi_{II}(x)=\frac{A}{\sqrt{|X'(x)|}}\sin\left[{X(x)\over \hbar}+ {\pi \over 4}\right]\ .
\end{equation} 		     

The approximate WKB expression for the wave function in the c.a.r. has the same form but with the classical reduced action $W_C(x)$ at the place of $X(x)$ and the classical momentum $p_C(x)$ instead of $X'(x)$.  This expression confirms that in the limit $\hbar\to 0$, the real part $X(x)$ of the quantum reduced action produces the classical one $W_C(x)$, while its derivative $X'(x)$ produces the classical momentum $p_C(x)$. In any case this directly follows from the WKB expansion.

The Eq. (25), which was already presented in [38] is a very interesting result of the present approach. Firstly, for the appropriate choice of the parameters, it exactly gives the wave function in the c.a.r. Moreover, it shows that every one-dimensional wave function can be represented in the same trigonometric form, by means of exact solutions of the QHJE. Finally, it proves that the real part $X(x)$ of the quantum characteristic function is a very important quantity, being the phase of the wave function, while its derivative controls the amplitude of this latter. It is worth to observe that while at the turning points the WKB expression diverges, the representation given by Eq. (25) does not present this problem, and exactly reproduces the wave function also there. The expression (25) also shows that $X'(x)$ has no zeros: indeed, these would correspond to poles in the wave function.

Eqs. (11) and (25) give a representation of the wave function in their respective regions, depending on 7 real parameters: the quantity A in Eq. (25), the $A_{I,III}$ in Eq. (11),  the complex constants $C_0$ in Eq. (20) and $C_1$ in Eq. (21). The matching of the expressions (11) and (25), is done by requiring the continuity of the wave function and its first derivative at the turning points, and by imposing the global normalization condition. However, these conditions leave some parameters indeterminate, which then have to be fixed arbitrarily. Indeed, just alike to the classical case, a constant can be added to any solution of Eq. (3), so we choose as first parameter to be fixed the value $W_G(x_1)$ in the left turning point, that will be usually put equal to 0. In particular, this implies $X(x_1) = 0$, in analogy with the WKB case, but a different choice will be done for the case of the hydrogen atom when $\ell=0$. In the classical limit, this value $X(x_1) = 0$ is the natural choice for a particle starting its motion from the left turning point, and proceeding in the positive sense of the x-axis. It is obviously possible to exchange the role of the two turning points, by imposing $X(x_2)= 0$, which  in the classical limit would describe a particle starting from the right turning point, and proceeding in the negative sense. But the more important condition on the real part $X(x)$ follows from the request that it has to be a continuous function, describing a periodic motion. The classical reduced action $W_C(x)$ is a continuous multi-valued function, and there are no limitations on its increment in one period of motion, but in the quantum case, the role of $X(x)$ as the phase of the wave function in Eq. (25) imposes a restriction. Indeed, according to the quantization rule Eq. (9), going from $x_1$ to $x_2$ the real part of the complex quantity $W_S(x)$ increases by $n\pi\hbar$, where $n$ is the number of poles of $p_{F,S}(x)$ (this can also be directly seen from the logarithmic terms in the explicit expressions for $W_S(x)$, in Sections 2 an 3). As a consequence of this and of the request of single valuedness of the wave function, we have to put the following condition on the second term in the r.h.s. of Eq. (22):

\begin{equation}
X_{add}(x_2)-X_{add}(x_1)=\left({\pi\over 2}\right)\hbar  \ .
\end{equation}      
In fact, in this case the increment of the function $X_{add}(x)$ in a period is $\pi\hbar$, so that the corresponding variation of the complete function $X(x)$, by effect of the variations of both the terms in the r.h.s. of Eq. (22) is $(n+1)\pi\hbar$ and the wave function in Eq. (24) after a period results simply multiplied by  +1 or -1, being its amplitude unchanged ($X'(x)$ is a single valued function). Eq. (26) fixes the values for the complex constant $C_0$, at the energy level given by the condition (9). Among the infinite family of solutions (20), (21) we will consider physically admissible only those with $C_0$ satisfying Eq. (26), or the solutions obtained by letting $|C_0|\to\infty$.

In this way we obtain from Eq. (22) a condition which is the generalization of the semiclassical quantization condition [10], in the form suitable for a semiperiod:
\begin{equation}
X(x_2)-X(x_1) = \int_{x_1}^{x_2} Re[p_{F,G}(x)] dx = \left(n+{1\over 2}\right)\pi\hbar\ .
\end{equation}   

Indeed, the semiclassical quantization condition is recovered from Eq. (27) in the semiclassical limit, because $X(x)\to W_C(x)$, and  $Re[p_{F,G}(x)] \to p_C(x)$ .
The requests above discussed fix the values of the complex constants $C_0$ and $C_1$ in Eqs. (20) and (21), while the other parameters are determined by imposing the continuity of the wave function and its first derivative at the turning points, and from the global normalization condition. The unicity theorem for differential equations applied to the SE then guarantees that the expressions (11) and (25), with the appropriate values of the parameters, exactly represent the wave function at the energy E in their respective region, by means of exact analytical solutions of the QHJE. This representation is possible only by using the general solutions.

When $|C_0| \to\infty$, the complex solution (20) reduces to the particular imaginary solution $p_{F,S} (x)$, the function $X(x)$ becomes the step function $Re[W_S(x)]$ increasing by $\pi\hbar$ at each pole of $p_{F,S}(x)$,  and it is no more possible to represents the wave function in the WKB form (25).

\section{THE HARMONIC OSCILLATOR: RESULTS AND DISCUSSION}
In this section we apply the general procedure exposed in Sect. 1, to the harmonic oscillator
$V(x) = {1\over 2} m \omega^2 x^2$.  Also for this system, the exact energy levels can be computed by means of the QHJE [13]. Moreover, Bhalla and coworkers [25] proved that the particular solution of the Riccati Eq. (7) for the n-th state of the harmonic oscillator is
\begin{equation}
p_{F,S}^{ho,n} (x) = i \left( m\omega x -{2n\sqrt{m\omega\hbar}H_{n-1}\left(\sqrt{m\omega\over\hbar}x\right) \over H_{n}\left( \sqrt{m\omega\over\hbar}x\right)	} \right)\ .
\end{equation}    
Here $H_n(x)$ is the n-order Hermite polynomial. This is the oscillator's quantum momentum function in the c.f.r.
By integrating this relation we get
\begin{equation}
W_{S}^{ho,n} (x) = i Y_{S}^{ho,n}(x) = i \left( {1\over 2} m\omega x^2 -\hbar \log \left[ H_{n}\left(\sqrt{m\omega\over\hbar}x\right)\right]	\right)
\end{equation}   
This is a particular solution of the QHJE, and gives the quantum reduced action in the c.f.r. It is immediate to verify that when it is inserted into Eq. (11), the correct wave function is obtained. As for the c.a.r., we need the general solutions so that, by inserting Eq. (28) into Eq. (20) according to the procedure exposed in Sect.1, and performing the integrations, we get:
\begin{equation}
p_{F,G}^{ho,n} (x) = p_{F,S}^{ho,n} (x)  + {e ^{m\omega x^2\over \hbar} \over H_{n}^2\left(\sqrt{m\omega\over\hbar}x\right) \left[{i\over \hbar }\int_0^x{{e ^{m\omega x^2\over \hbar} \over  H_{n}^2\left(\sqrt{m\omega\over\hbar}x\right)}}dx + C_0^{ho,n}\right]} \ .
\end{equation}      
For each value of the integer n, the integral in the second term of the r.h.s. of this equation can be expressed by means of elementary functions and the error function of imaginary argument, which is connected to the Dawson integral [40].
The constants $C_0^{ho,n}$, as obtained from the condition (26), are purely real, and are given by :
 \begin{equation}
 C_0^{ho,n}={1\over \hbar} \int_0^{x_2}{e ^{m\omega x^2\over \hbar} \over  H_{n}^2\left(\sqrt{m\omega\over\hbar}x\right)}dx
 \end{equation}			
Some numerical results for $p_{F,G} (x)$ are reported in Figs. 1 and 2.

\begin{figure}[!h]
\includegraphics[scale=1]{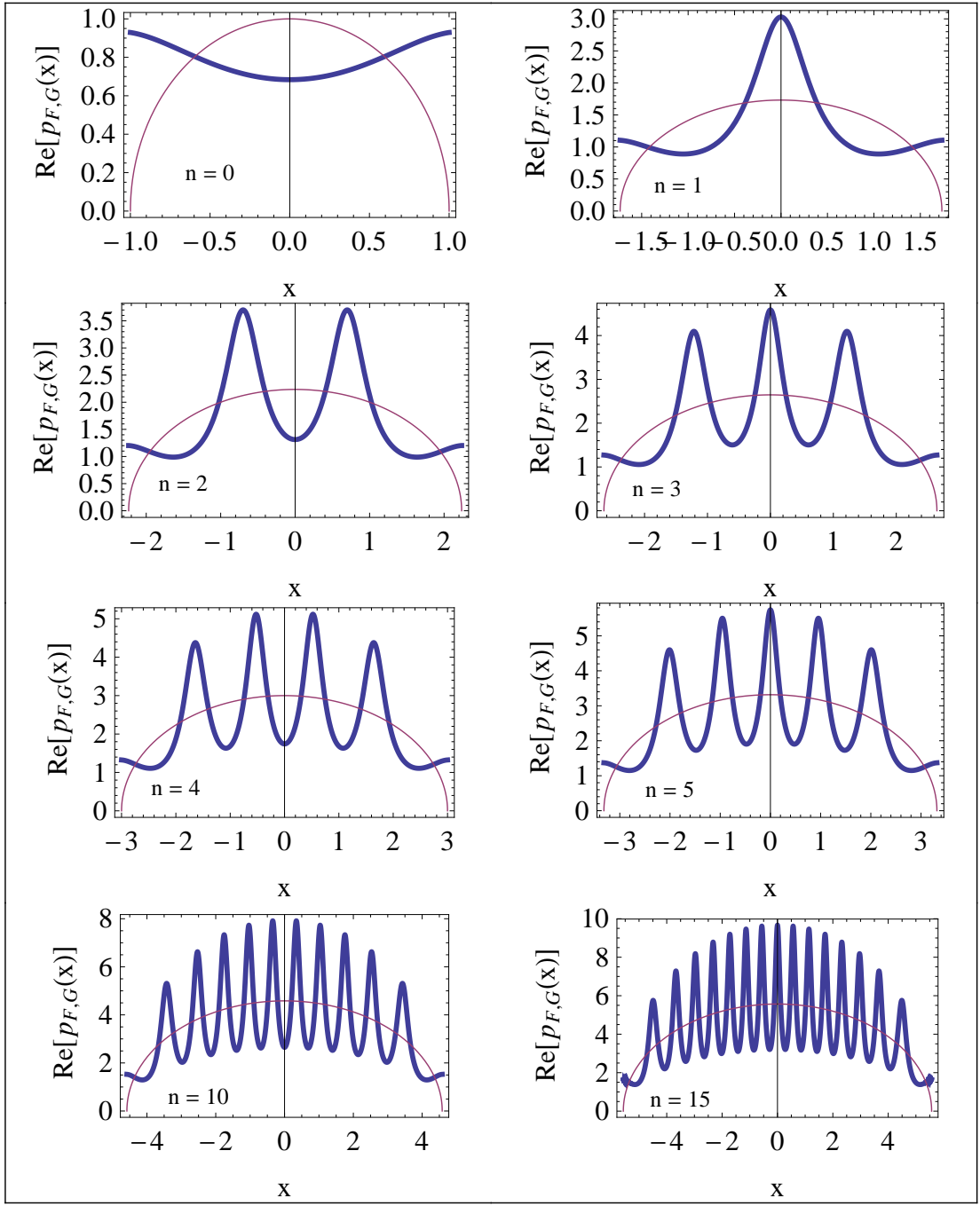}%
\caption{The real part $Re[p_{F,G}^{ho,n} (x) ]$  of the quantum momentum function for the harmonic oscillator is reported for  various values of $n$ (thick lines). For comparison, the corresponding classical momentum $p_C(x)$ is also plotted (thin lines).}
\end{figure} 

In Fig. 1 the real part $Re[p_{F,G}^{ho,n} (x) ]$ of the quantum momentum function for some values of $n$ from $n=0$ to $n=15$ is reported (thick lines). For comparison, the corresponding classical momentum $p_C(x)$ is also plotted (thin lines). The real part of $p_{F,G}^{ho,n} (x)$  comes only from the second term in the r.h.s. of Eq. (30), which has a real part due to the presence of the constant $C_0$. As seen from the figures, $Re[p_{F,G}^{ho,n} (x) ]$ is an even function of $x$, always positive, and presents $n$ peaks of finite heights near or exactly in correspondence of the zeroes of the $n$-th Hermite polynomial. The peaks' height is proportional to $n^{1\over 2}$ so, while increasing the value of n, the peaks become higher and sharper, and their centers tend to follow more closely the profile of the classical momentum $p_C(x)$;  differently from this latter however, $Re[p_{F,G}^{ho,n} (x) ]$ has a value different from zero at the turning points.

\begin{figure}[!h]
\includegraphics[scale=1]{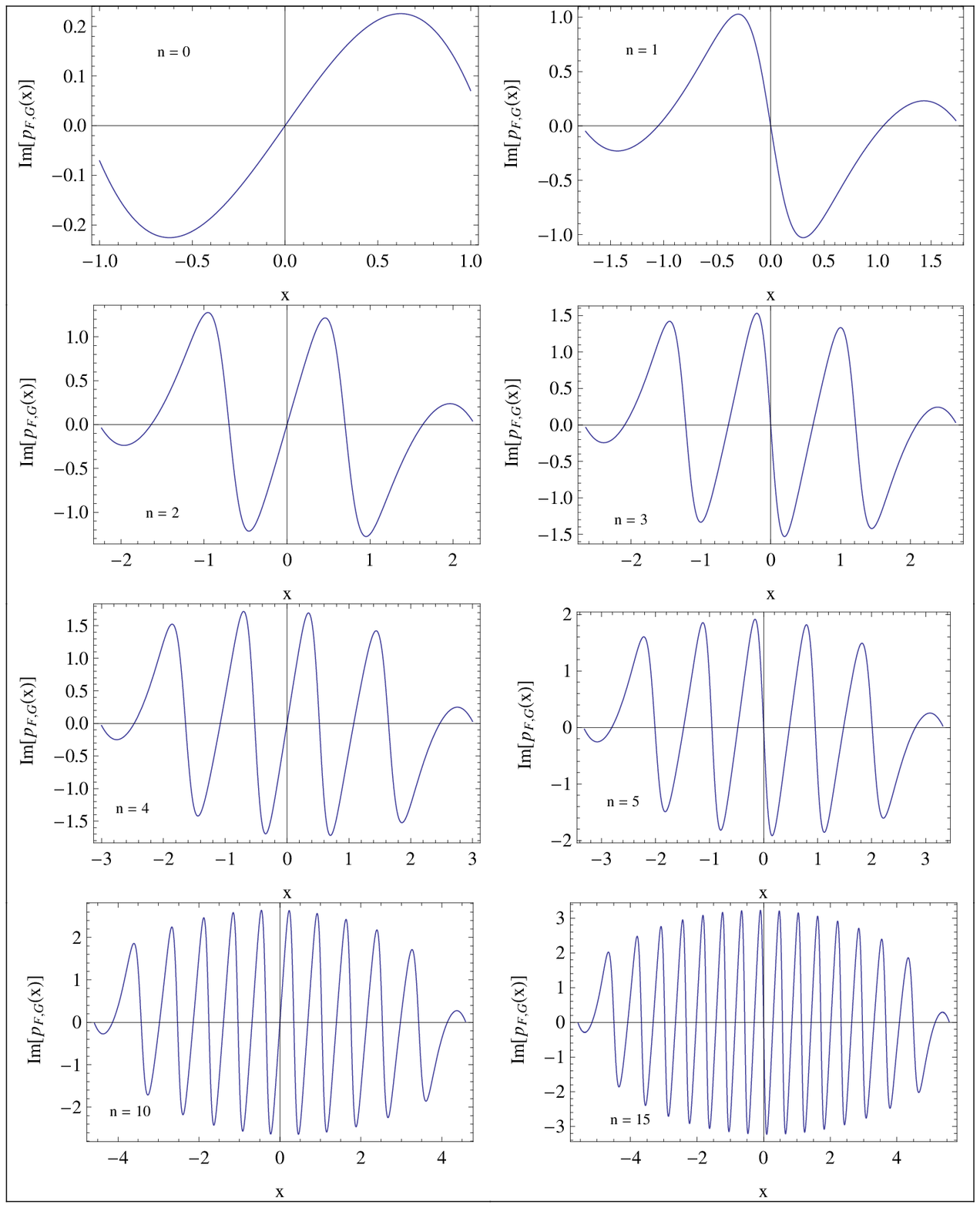}
\caption{The imaginary part $Im[p_{F,G}^{ho,n} (x) ]$ of the quantum momentum function for the harmonic oscillator is reported for  various values of $n$. }
\end{figure}   

In Fig. 2 the imaginary part $Im[p_{F,G}^{ho,n} (x) ]$ of the quantum momentum function for the same values of $n$ is plotted. As seen from the figures, this function presents oscillations of finite amplitude around zero, as opposed to the poles  of $Im[p_{F,S}(x)]$; the number of oscillations increases with $n$. These oscillations are the result of the competition between the opposite poles of two terms in the r.h.s. of Eq. (30), in correspondence of the nodes of the $n$-th Hermite polynomial.
By integrating Eq. (30), we get the oscillator's quantum reduced action in the c.a.r.:
\begin{equation}
W_{G}^{ho,n} (x) = W_{S}^{ho,n}(x) +{\hbar\over i} \log \left[ {i\over \hbar} \int_0^x{{e ^{m\omega x^2\over \hbar} \over  H_{n}^2\left(\sqrt{m\omega\over\hbar}x\right)}}dx + C_0^{ho,n}\right]+ C_1^{ho,n}  \ .
\end{equation}   
Given $C_0^{ho,n}$ , the complex constants $C_1^{ho,n}$ are determined according to Sect. 1, by requiring that $W_{G}^{ho,n} (x_1)  = 0$.
The real part of this expression is:
\begin{equation}
X^{ho,n} (x) = \hbar\ Arg \left[H_{n}\left(\sqrt{m\omega\over\hbar}x\right)\right] +\hbar\ Arg \left[ {i\over \hbar} \int_0^x{{e ^{m\omega x^2\over \hbar} \over  H_{n}^2\left(\sqrt{m\omega\over\hbar}x\right)}}dx + C_0^{ho,n}\right]+ Re[C_1^{ho,n}] \ .
\end{equation}   

In the following, the first term in the r.h.s. will be denoted as $X_S^{ho,n} (x)$ and the second term as $X_{add}^{ho,n}(x)$. For $n=0$, the $Arg$ in $X_S^{ho,n} (x)$ is zero, and when $n$ is different from zero it varies by $\pm\pi ( mod\ 2\pi)$ at each passage of $x$ through a node of the Hermite polynomial. For $n=0$,  $X_{add}^{ho,0}(x)$ is a continuous function, starting from $-\pi\hbar/4$ in $x_1$ and ending in $x_2$ with the value $\pi\hbar/4$, so that its total variation is $\pi\hbar/2$. $X^{ho,0} (x)$ so reduces to $X_{add}^{ho,0}(x)+ Re[C^{ho,0}_1]$. Its graph is reported in the left upper box of Fig. 4.

\begin{figure}[!h]
\includegraphics[scale=1]{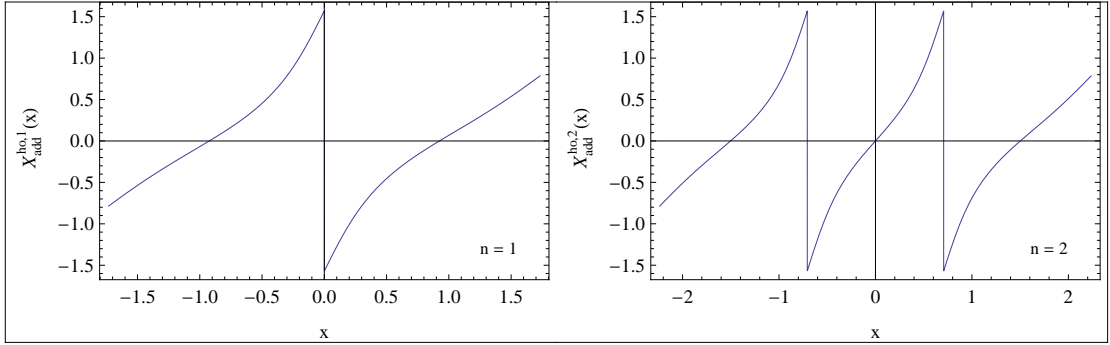}
\caption{The graphs of $X_{add}^{ho,1}(x)$ (left box)  and $X_{add}^{ho,2}(x)$ (right box).}
\end{figure}     

The graphs of $X_{add}^{ho,1}(x)$  and $X_{add}^{ho,2}(x)$  are reported in Fig. 3. The function  $X_{add}^{ho,1}(x)$  starts from the value $-\pi\hbar/4$ in $x_1$, has a jump discontinuity in the origin, which is the node of the first order Hermite polynomial, falling from the value  $\pi\hbar/2$ to $- \pi\hbar/2$, and then increases again and ends in $x_2$ with the value $\pi\hbar/4$. This jump of $X_{add}^{ho,1}(x)$  is compensated by the opposite jump of the first term $X_{S}^{ho,1}(x)$ in Eq.(33), which increases by $\pi\hbar$ passing throughout the origin.  The total variation of $X^{ho,1}(x)$ passing from $x_1$ to $x_2$ is ${3\pi\hbar\over 4} +{3\pi\hbar\over 4}={3\pi\hbar \over 2} = (1+ {1\over 2}) \pi\hbar$, and $X^{ho,1}(x)$ is a continuous increasing function, as the classical characteristic function $W_C(x)$.  For $n=2$, $X^{ho,2}(x)$ has two such jumps at the nodes of the 2-th Hermite polynomial, and between them $X_{add}^{ho,2}(x)$ increases by $\pi\hbar$, passing from -- $\pi\hbar/2$ to  $\pi\hbar/2$. The two singularities are compensated by the two opposite variations, each of $\pi\hbar$, of the first term $X_{s}^{ho,2}(x)$ in Eq. (33), if we admit that the Arg in it is not restricted to the principal value, but increases by $\pi$ at each changing of sign of the Hermite polynomial. The total variation therefore is $(2+ {1\over 2}) \pi\hbar$. For arbitrary $n$, these considerations can be repeated for each pair of consecutive nodes of the $n$-th Hermite polynomial, so that $X^{ho,n}(x)$ is a continuous increasing function, whose total variation passing from $x_1$ to $x_2$ is $(n+ {1\over 2}) \pi\hbar$.  The first term in the r.h.s. of Eq. (33) alone would give a staircase increasing function, while the second term smooths this behavior.

\begin{figure}[!h]
\includegraphics[scale=1]{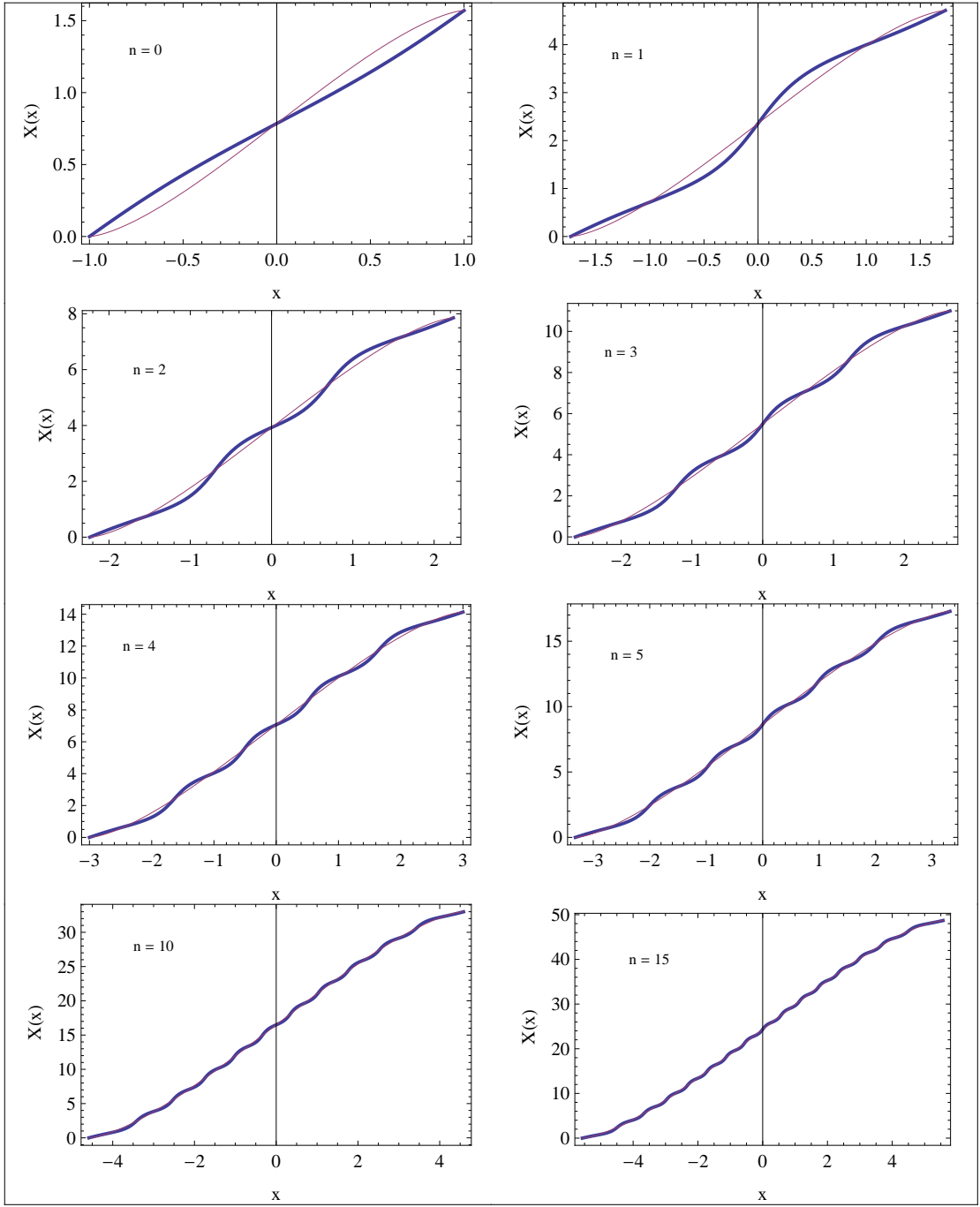}
\caption{The real part $X^{ho,n} (x)$ of the quantum reduced action for the harmonic oscillator is plotted for various values of the quantum number $n$, from $n=0$ to $n=15$, as computed from Eq. (32) (thick lines) together with the corresponding classical quantity $W_C(x)$ at the same energy (thin lines).}
\end{figure}           

Fig. 4 reports the graphs of the real part $X^{ho,n} (x)$ of the quantum reduced action, for various values of n, from n=0 to n=15, as computed from Eq. (33) (thick lines) together with the corresponding classical quantity $W_C(x)$ at the same energy (thin lines). As seen from the figures, $X^{ho,n}(x)$ follows the profile of the classical corresponding function $W_C (x)$, waving so that their difference is an oscillating function, with a maximum amplitude independent on $n$, while the number of oscillations increases with $n$. The ripples of $X^{ho,n}(x)$ cause the peaks in $Re[p^{ho,n}_{F,G},(x)]$. The imaginary parts $Y^{ho,n}(x)$ of the quantum reduced action are not reported in the figures, as their behavior can be easily deduced by means of Eq. (15) from the graphs in Fig. 2: these are oscillating functions, similar to $Re[p^{ho,n}_{F,G}(x)]$.
\begin{figure}[!h]
\includegraphics[scale=1]{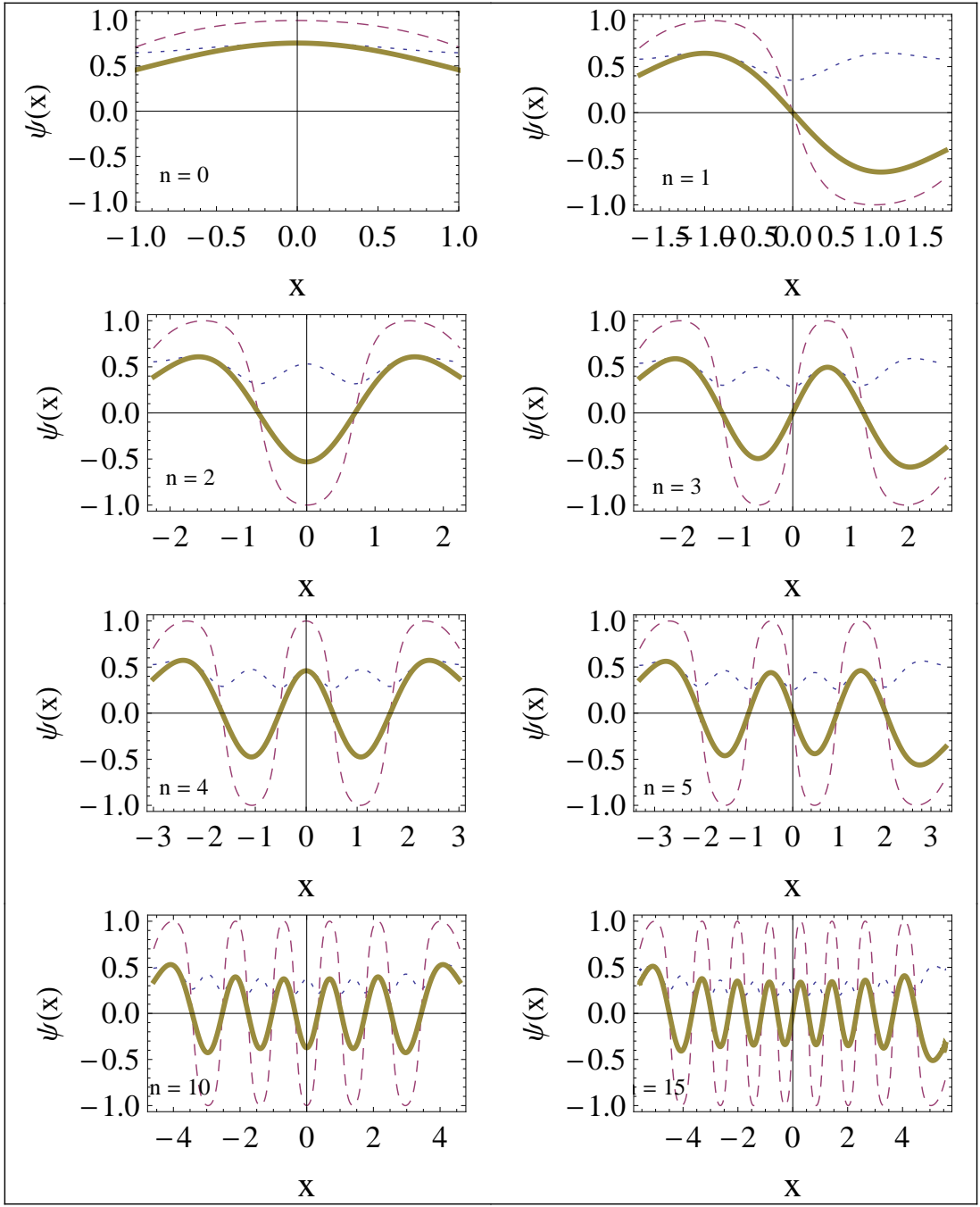}
\caption{For the harmonic oscillator, the functions $\sin[X^{ho,n}(x) / \hbar + \pi /4]$ (dashed lines), $1/\surd |X^{ho,n'}(x)|$ (dotted lines) , and finally their product (thick lines)  which according to Eq. (25) exactly gives the wave function in the c.a.r., are reported for various values of the quantum number n.}
\end{figure}  
The expression for $X^{ho,n}(x)$ so obtained has to be inserted in Eq. (25) in order to get the exact sinusoidal representation of the wave function in the classically allowed region.

Fig. 5 presents the functions $\sin[X^{ho,n}(x) / \hbar + \pi /4]$ (dashed lines) , $1/\surd |X^{ho,n'}(x)|$ (dotted lines), and finally their product (thick lines) which according to Eq. (25) exactly reproduces the well known oscillator's wave function in the c.a.r. In all the computations, the values $\hbar = 1$ for the Planck's constant, $m=1$ and $\omega =1$ for the particle's mass and angular frequency were used.

From Eq. (25) and Figs. 4 and 5 it is clear that with our choice $X^{ho,n}(x_1)=0$, the phase of the wave function in Eq. (25) starts from $\pi/4$  at the left turning point and monotonically increases with $x$.  Each passage of $X^{ho,n}(x)$ through a value $(k - 1/4)\pi\hbar$  where $k$ is an integer gives a node of the wave function, and the oscillations of this latter follow those of the factor $\sin[X^{ho,n}(x)+ \pi/4 ]$. Incidentally, it is interesting to see that the zeroes of the Hermite polynomials are tied to the values of a particular solution of a third order nonlinear equation. The amplitude of the wave function is instead regulated by the derivative ${X^{ho,n}}'(x)$, through the factor $1/\surd[{X^{ho,n}}'(x)]$. In conclusion, the wave function is completely governed by the real part $X^{ho,n}(x)$ of the quantum reduced function, which appears so to be a fundamental quantity in quantum mechanics.

\section{THE HYDROGEN ATOM}
The method described in Sect. 1 can be applied to separable three dimensional cases, whose most important example is the hydrogen atom. Indeed, as well known [10], for central potentials $U(\rho)$ where $\rho$ is the distance from the origin, and by denoting with $R(\rho)$ the radial part of the factorized wave function, one obtains the following equation for the function $f(\rho) = \rho R (\rho)$:
\begin{equation}
{\hbar\over 2 m}  {d^2 f\over d\rho^2} + \left[ E-U(\rho)-{l(l +1)\over 2m\rho^2}\right]  f=0\ ,
\end{equation}  
 This equation is formally the SE for one dimensional motion of a particle in a potential
\begin{equation}
V(\rho) = U(\rho) + {l(l +1)\over 2m\rho^2}\ .
\end{equation}  
Therefore, the procedure of Sect.2 can be applied to the radial part of the motion, leaving untouched the angular part of the factorized wave function. Let us consider in particular the case of the Kepler problem for the electron in the hydrogen atom, with the potential $U(\rho) = - \alpha/ \rho$. It is easy to see that for the state with radial quantum number $n$ and orbital quantum number $l$, the particular solution of the Riccati equation (7) for the hydrogen atom is:
\begin{equation}
p^{k,n,l}_{F,S} (r) =-i{\hbar\over a_0}\left(-{1\over n}+{l +1\over r}+{2L^{2l+2}_{n-l-2} (2r/n) \over L^{2l+1}_{n-l-1}(2r/n)}\right)\ ,
\end{equation}  
where $a_0$ is the Bohr's radius, $r$ is the adimensional variable $r=\rho/ a_0$, and the apex "k" stands for Kepler problem.
By integrating with respect to $\rho$ we get:
\begin{equation}
W_S^{k,n,l}[r] = {i\hbar r\over n}- i\hbar (l+1) \log [r] - i\hbar\log [L^{2l+1}_{n-l-1}(2r/n)]\ .
\end{equation}			

These are the particular solutions, giving the quantum momentum function and the quantum reduced action in the c.f.r., respectively. The general solutions  $p_{F,G}^{ k,n,l}$ is:
\begin{equation}
p_{G,S}^{k,n,l} (r) = p_{F,S}^{k,n,l} (r)  +{ {{e ^{2r\over n} r^{-2(1+l)}}\over {\left[L^{2l+1}_{n-l-1} (2r/n)\right] ^2}} \over \int_0^r \left[{{e ^{2r\over n} r^{-2(1+l)}}\over\left[L^{2l+1}_{n-l-1} (2r/n)\right] ^2} \right] dr + C_0^{k,n,l} } \ .
\end{equation}            
For each value of the integer numbers $n$ and $l$, the integral in this expression can be calculated in terms of elementary functions and the Exponential Integral function [41].

Integration of the last equation gives:
\begin{equation}
W_{G}^{k,n,l} (r) = W_{S}^{k,n,l} (r)  + ({\hbar\over i})\log \left[ \left({i\over \hbar}\right) \int_0^r {e ^{2r\over n} r^{-2(1+l)}\over\left[L^{2l+1}_{n-l-1} (2r/n)\right] ^2} dr + C_0^{k,n,l}\right] + C_1^{k,n,l}\ .
\end{equation}            
Its real part $X^{k,n,l}(r)$ is given by:
\begin{equation}
X^{k,n,l} (r) = Re  \left[ W_{S}^{k,n,l} (r) \right] + \hbar Arg \left[ \left({i\over \hbar}\right)\int_0^r {e ^{2r\over n} r^{-2(1+l)}\over\left[L^{2l+1}_{n-l-1} (2r/n)\right] ^2} dr + C_0^{k,n,l}\right] + Re[C_1^{k,n,l}]\ .
\end{equation}            
The constants $C_0^{k,n,l}$  and $C_1^{k,n,l}$  are computed as described in Sect. 1 and for the hydrogen atom are complex numbers. Due to the singularity of the potential at $r=0$, for their determination we need to treat differently the cases with $l \neq 0$, and $l=0$.
Let us define
\begin{equation}
g_{n,l}(r)=\int_0^r\left[{e ^{2r\over n} r^{-2(1+l)}\over\left[L^{2l+1}_{n-l-1} (2r/n)\right] ^2}\right] dr \ .
\end{equation} 	
When $l\ne 0$, the constants  $C_0^{k,n,l}$  satisfying Eq. (26) are given by:
\begin{equation}
C_0^{k,n,l} = {g_{n,l}(r_2) - g_{n,l}(r_1) \over 2} -i {g_{n,l}(r_2) + g_{n,l}(r_1) \over 2} \ ,
\end{equation} 	
where $r_1$ and $r_2$ are the left and right turning points, respectively.

Given $C^{k,n,l}_0$, the complex constant $C^{k,n,l}_1$  is obtained as usual. The numerical results are presented in the figures 6-9.

\begin{figure}[!htbp]
\includegraphics[scale=1]{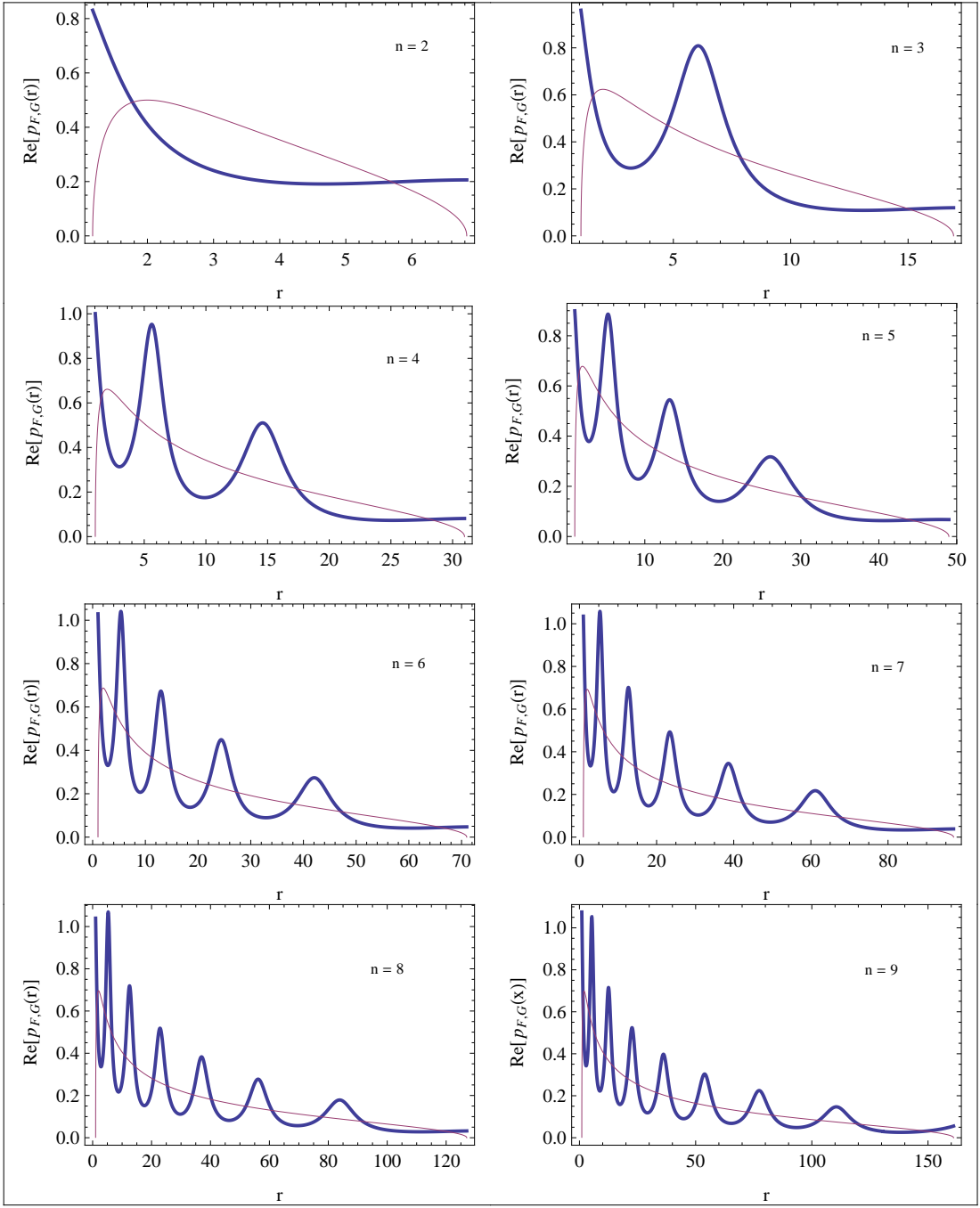}
\caption{The real part $Re[ p_{F,G} (r)]$ of the Kepler's quantum momentum function, for the radial quantum number $n$ going from 2 to 9, and for the value $l=1$ of the orbital number,  as computed from Eq. (37) (thick lines) together with the corresponding classical quantity $p_C(r)$ (thin lines)}
\end{figure}    

Fig. 6 reports the real part $Re[p^{k,n,l}_{F,G}(x)]$ of the quantum momentum function of the Kepler problem for the radial quantum number n going from 2 to 9, and for the value $l=1$ of the orbital number,  as computed from Eq. (38) (thick lines), together with the corresponding classical quantity $p_C(r)$ (thin lines). The corresponding imaginary part $Im[p^{k,n,l}_{F,G}(r)]$  is reported in  Fig. 7.
\begin{figure}[!htbp]
\includegraphics[scale=1]{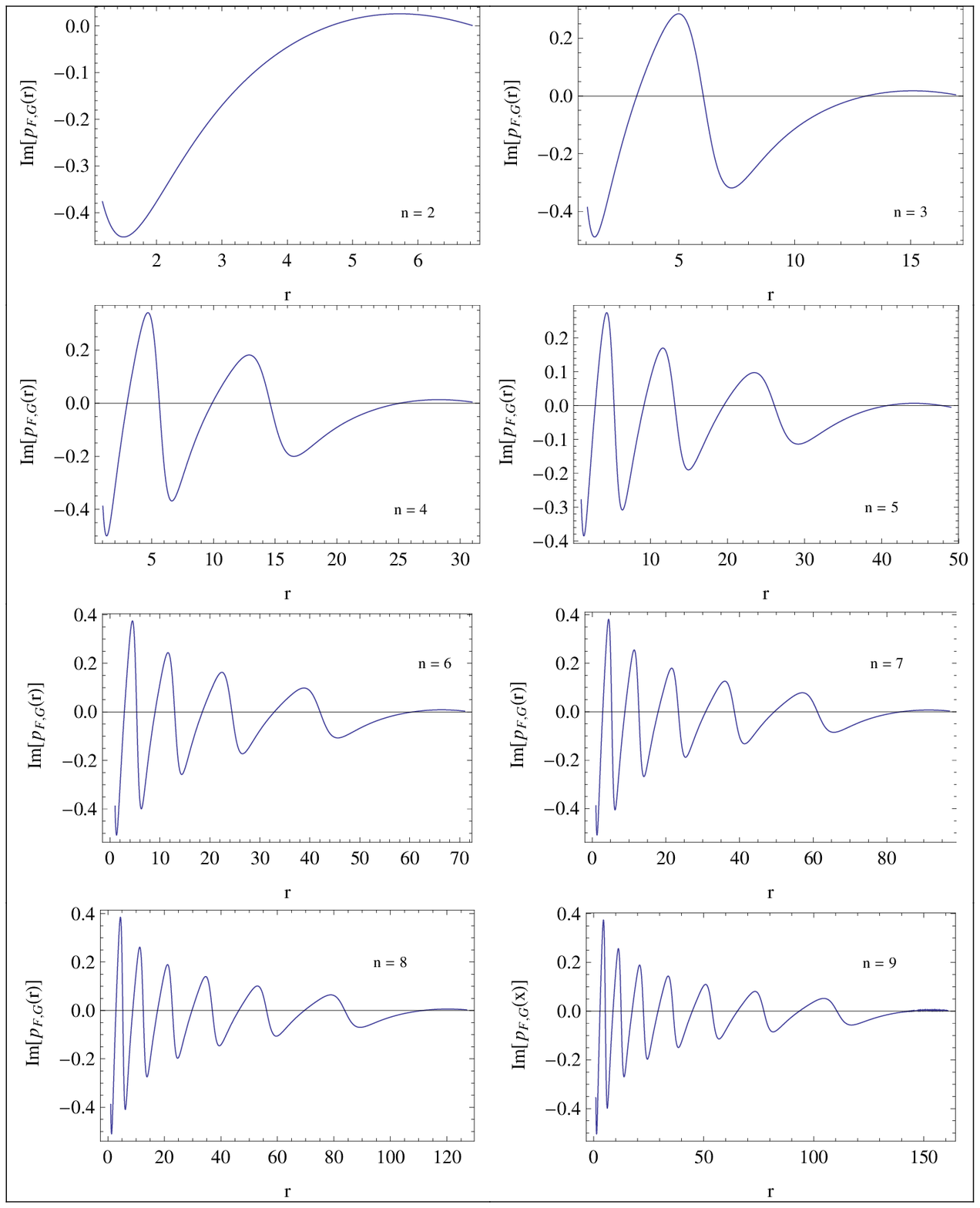}
\caption{The imaginary part $Im[ p_{F,G} (r)]$ of the quantum reduced action of the Kepler problem, for the radial quantum number n going from 2 to 9, and for the value $l=1$ of the orbital number.}
\end{figure}     
For the same values of $n$ and $l$, the real part of the quantum reduced action $X^{k,n,l}(r)$ is reported in Fig. 8 (thick lines), while the thin lines refer to the corresponding classical reduced action $W_C(r)$. The results are similar to those for the harmonic oscillator. $Re[p^{k,n,l}_{F,G}(r)]$, which comes from the second term in Eq.(38) differently from $p_C(r)$ is always positive and has $n-l+1$ peaks, in correspondence of the nodes of the Laguerre polynomial. While increasing n, the centers of the peaks tend to follow more closely the profile of the classical momentum. $Im[p^{k,n,l}_{  F,G}(x)]$ instead is an oscillating function, with a number of oscillations increasing with $n$; their amplitude decreases while increasing  $r$.
\begin{figure}[!htbp]
\includegraphics[scale=1]{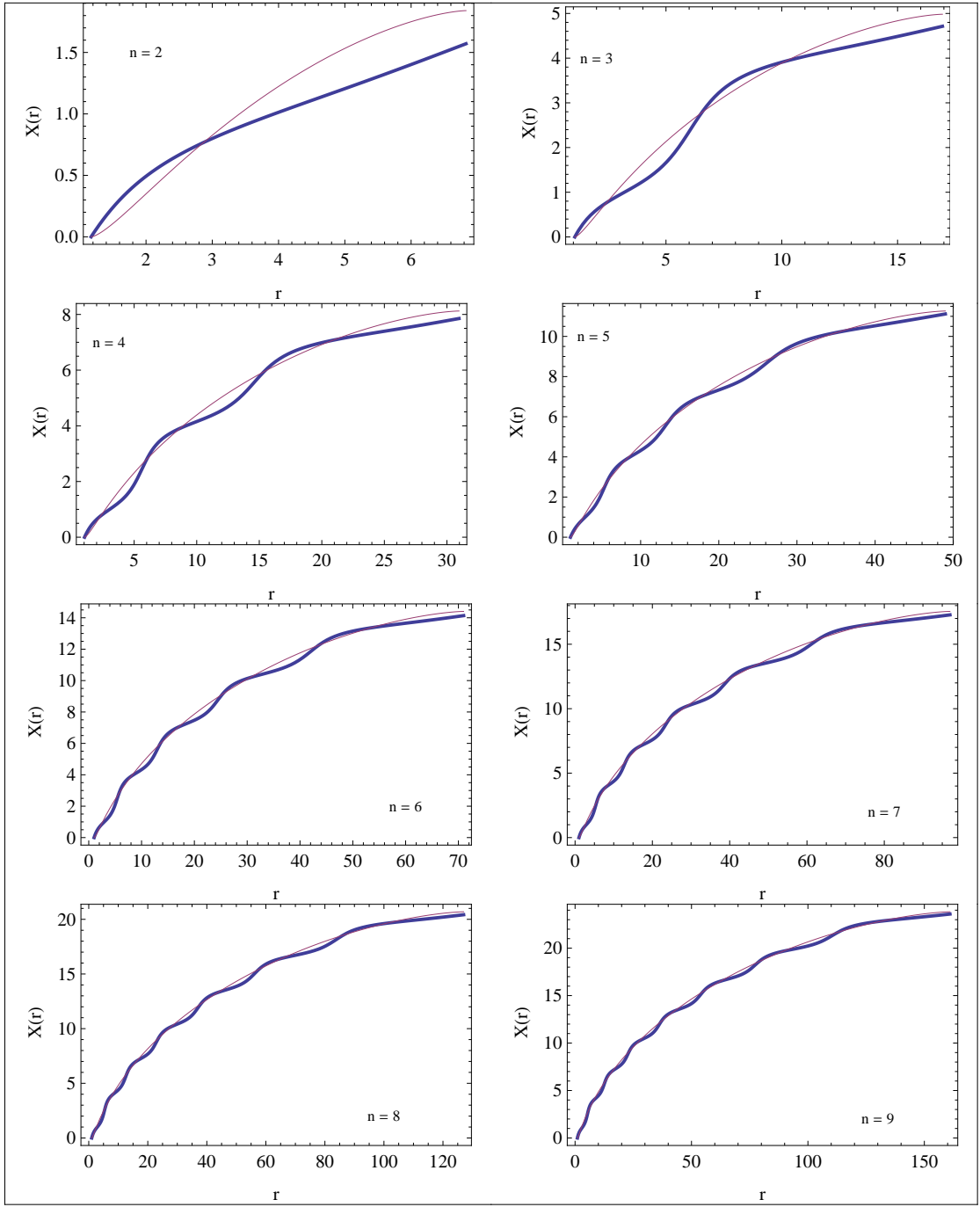}
\caption{The real part $X^{k,n,l} (x)$ of the quantum reduced action of the Kepler problem, for the radial quantum number n going from 2 to 9, and for the value $l=1$ of the orbital number,  as computed from Eq. (39) (thick lines) together with the corresponding classical quantity $W_C(r)$ (thin lines)}
\end{figure}  
Also for the Kepler problem, the real part $X^{k,n,l} (r)$ of the quantum reduced action is a waving function, whose profile tends to follow closer and closer the profile of the classical reduced action while increasing $n$. Its ripples causes the peaks in the real part of the quantum momentum function.

In Fig. 9 are reported the functions $\sin[X^{k,n,l}(r) / \hbar + \pi /4]$ (dashed lines), $1/\surd |X^{k,n,l'}(r)|$ (dotted lines), and finally their product (thick lines) which according to Eq. (25) exactly gives the function $f(r)$ which is strictly related to the radial part of the factorized wave function. In the computation the values $\alpha=1$, $\hbar=1$, $m=1$ were used.

\begin{figure}[!htbp]
\includegraphics[scale=1]{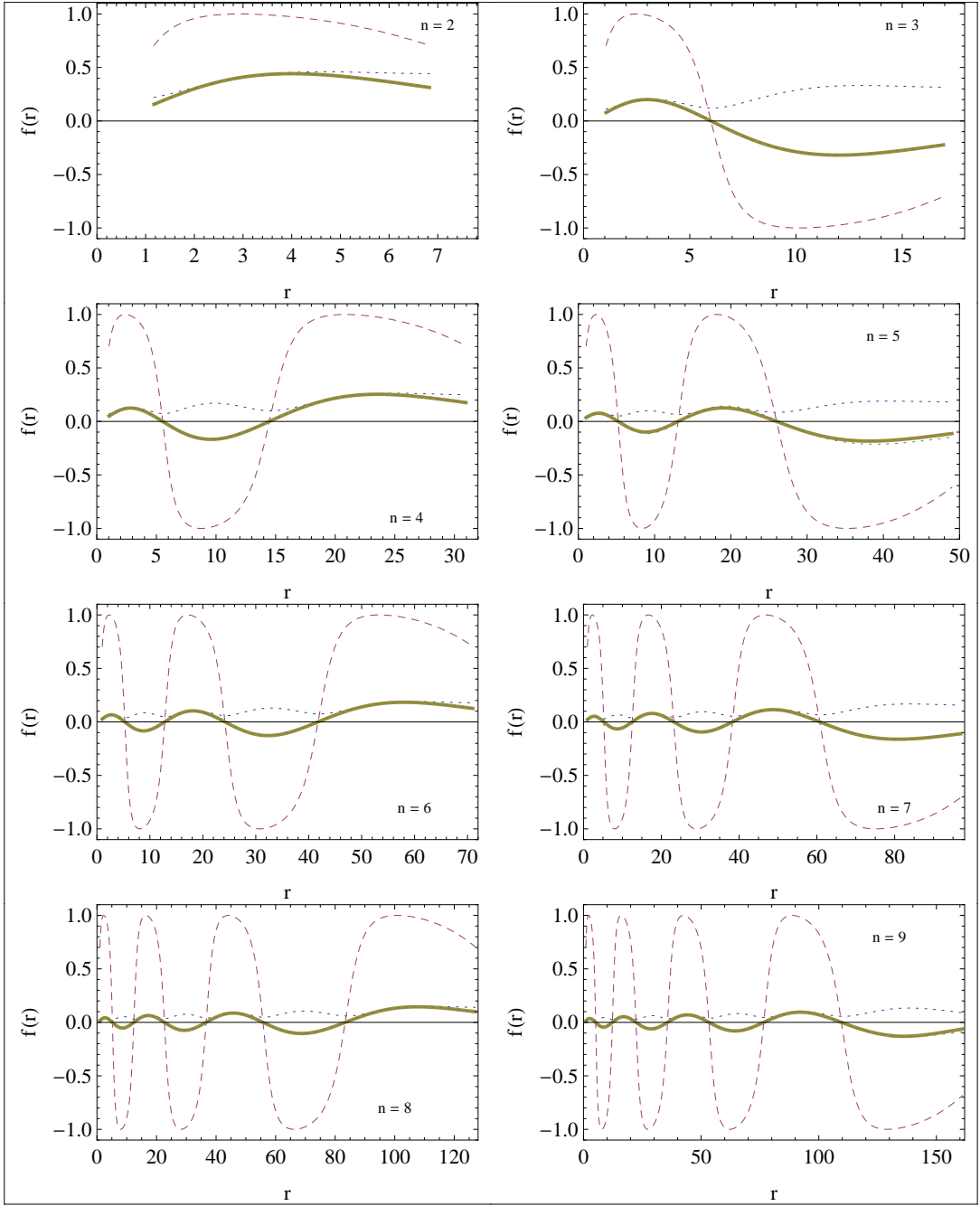}
\caption{The functions $\sin[X^{k,n,l}(r) / \hbar + \pi /4]$ (dashed lines) , $1/\surd |X^{k,n,l'}(x)|$ (dotted lines), and finally their product (thick lines) which according to Eq. (25) exactly gives the function $f(r)$ which is related to the radial part of the factorized wave function.}
\end{figure}

\begin{figure}[!htbp]
\includegraphics[scale=1]{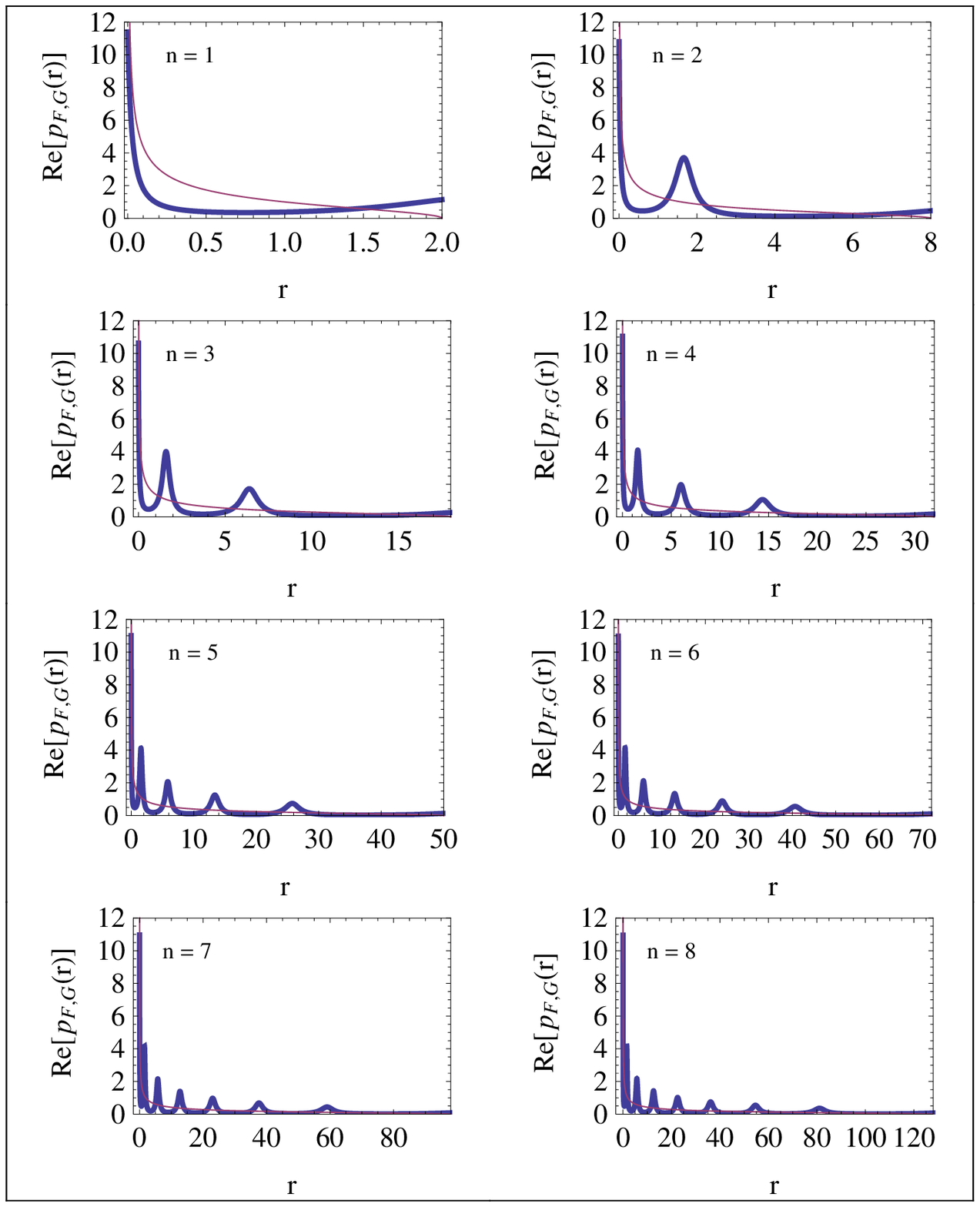}
\caption{The real part $Re[ p_{F,G} (r)]$ of the Kepler's quantum momentum function, for the radial quantum number $n$ going from 2 to 9, and for the value $l=0$ of the orbital number,  as computed from Eq. (37) (thick lines) together with the corresponding classical quantity $p_C(r)$ (thin lines).}
\end{figure} 

\begin{figure}[!htbp]
\includegraphics[scale=1]{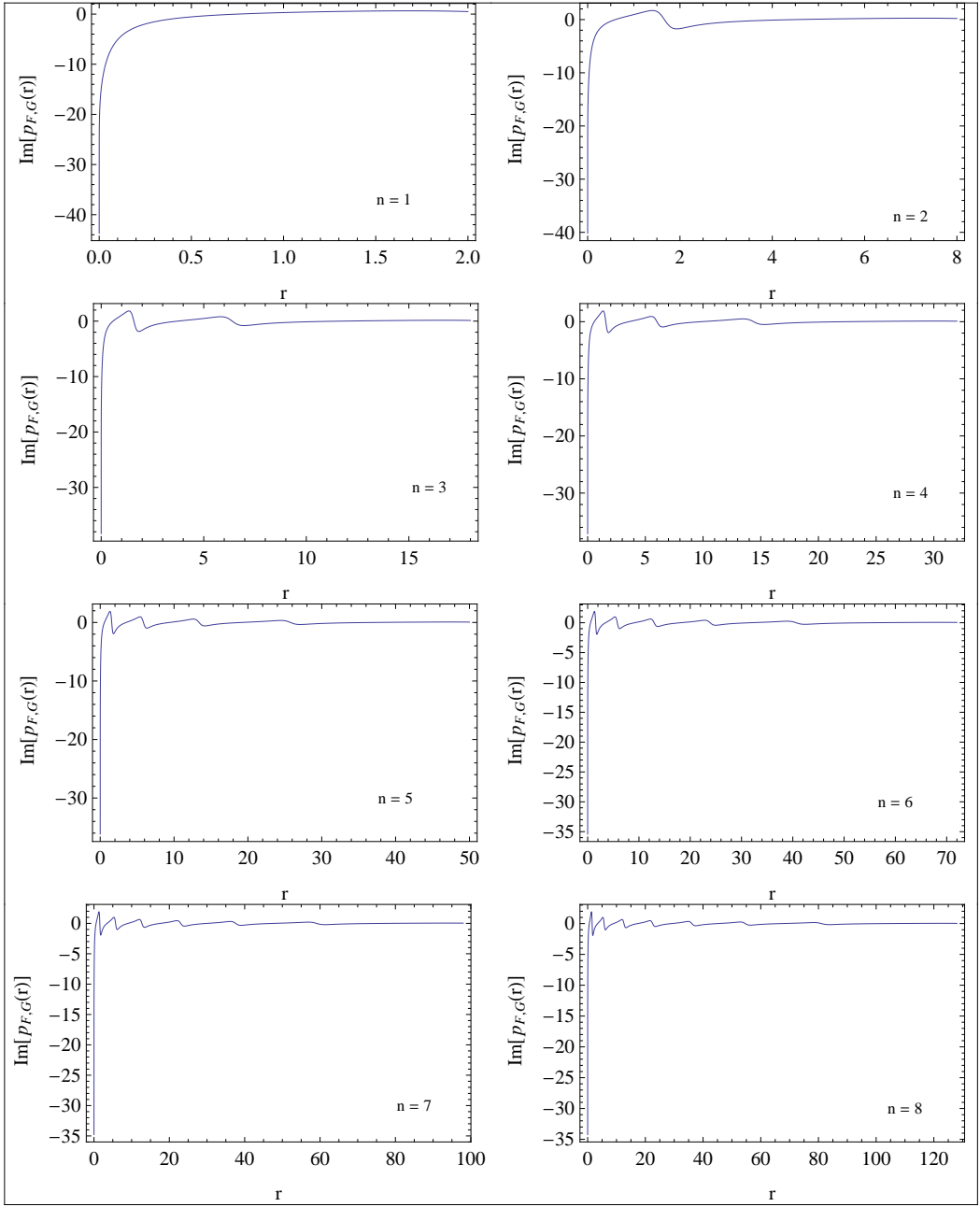}
\caption{The imaginary part $Im[p_{F,G}(r)]$ of the Kepler's quantum momentum function, for the radial quantum number $n$ going from 2 to 9, and for the value $l=0$ of the orbital number.}
\end{figure}

\begin{figure}[!htbp]
\includegraphics[scale=1]{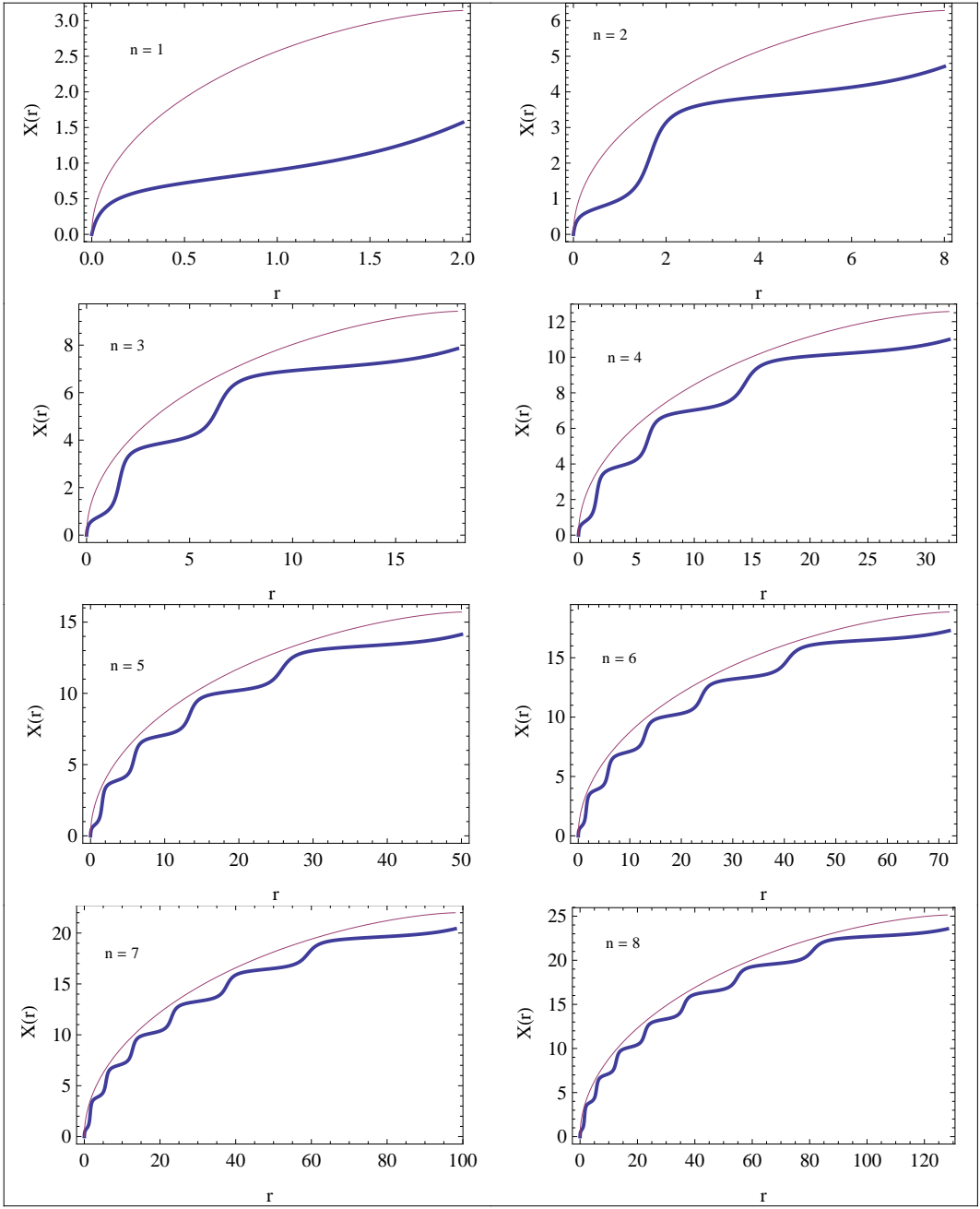}
\caption{The real part $X^{k,n,0} (r)$ of the Kepler's quantum reduced action, for the radial quantum number $n$ going from 2 to 9, and for the value $l=0$ of the orbital number,  as computed from Eq. (40) (thick lines) together with the corresponding classical quantity $p_C(r)$ (thin lines)}
\end{figure}

\begin{figure}[!htbp]
\includegraphics[scale=1]{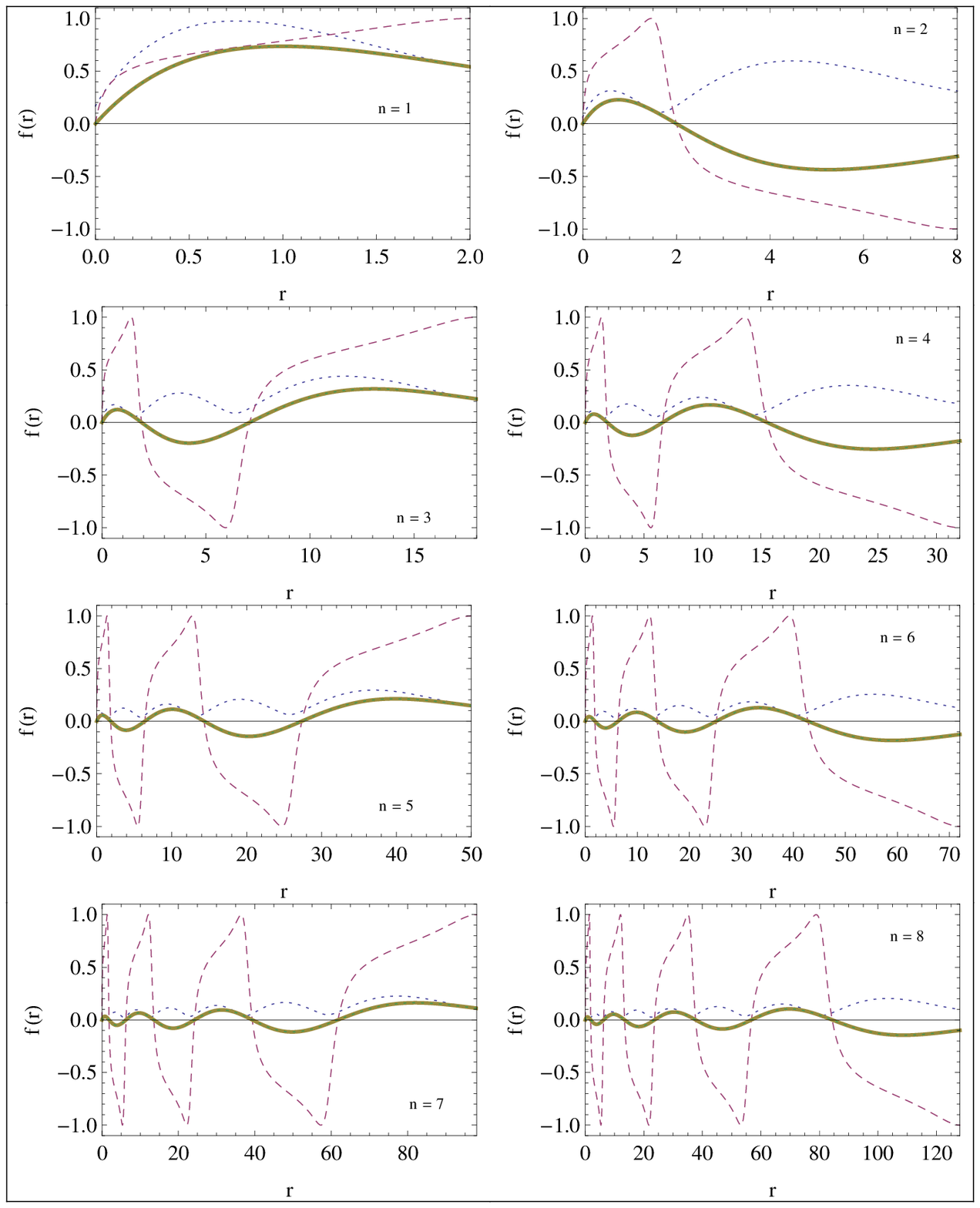}
\caption{For the Kepler's problem with $l = 0$, the functions $\sin[X^{k,n,0}(x) / \hbar + \pi /4]$ (dashed lines) , $1/\surd |{X^{k,n,0}}'(r)|$ (dotted lines), and finally their product (continuous lines) which according to Eq. (25) exactly gives the function $f(r)$ which is strictly related to the radial part of the factorized wave function. }
\end{figure}

The cases with $l=0$ can be treated in the same way, but now the left turning point is the origin $r=0$ itself, because also this point is reachable for the particle. In order to continue to use the same WKB-like expression (25) for the wave function, we choose then

\begin{equation}
X^{k,n,0}(0)= -{\pi\over 4}\hbar
\end{equation}             

It is easy to see that in this case the condition (26) fixes only the imaginary part of the  constants
$C^{k,n,0}_0$, as
\begin{equation}
Im[C^{k,n,0}_0]  = - g(r_2)\ ,
\end{equation}					
while the real part can be chosen arbitrarily. In the numerical computation reported in the figures
the following values were used:
\begin{equation}
Re[C^{k,n,0}_0] = g(r_2)
\end{equation}  

The corresponding graphs are given in the figures 10-13. The results are similar to those for the case $l\ne 0$, apart for the fact that the function $f(r)$ has now a node in the origin.

As happens for the harmonic oscillator, the real part $X^{k,n,0}(r)$ of the quantum characteristic function follows waving the classical reduced action at the same energy. Its ripples causes peaks in the real part of the quantum momentum function $Re[ p_{F,G}^{k,n,0} (r)]$. The number of peaks is $n-l+1$, i.e. the number of nodes of the radial function. The graph for $Re[ p_{F,G}^{k,n,0} (r)]$ tends to follow the graph for the corresponding classical momentum $p_C(r)$, but differently from this latter, which diverges for $r=0$, has a finite value there.

\section{THE CLASSICAL LIMIT}
When $\hbar \to 0$, the QHJE reduces to the classical HJE, so that the quantum characteristic function $W(x,E)$ has to tend in a way to be investigated to the classical one $W_C(x,E)$.

In order to examinate more closely what happens in the classical limit, it is necessary to specify the potential and distinguish between the cases of c.f.r. and of c.a.r.

In the c.f.r., the limit for $\hbar\to 0$ of the particular imaginary solutions of the Riccati equation $p_{F,S}(x)$ for the harmonic oscillator and the hydrogen atom is immediate to calculate, as can be seen from the explicit expressions given in Eqs. (28) and (36), respectively, and gives the classical, imaginary  quantities  $W_C (x)$ and $p_C(x)$ in these regions.

For the c.a.r., one should investigate the limit by means of the general solutions given by Eqs. (20), (21), (36) and (37). However, due to the non-analytical dependence of the various functions on $\hbar$, it is not possible to expand in power series of this quantity; nevertheless, a clear indication of what happens in the limit can be obtained by means of numerical computations. These have to be done at fixed energy (that in particular implies that the turning points do not change) and by comparing the various functions computed for increasing values of $n$, which corresponds to repeat the computations with decreasing values of the Planck's constant. Some results for the harmonic oscillator are reported in Fig. 14.

\begin{figure}[!h]
\includegraphics[scale=0.8]{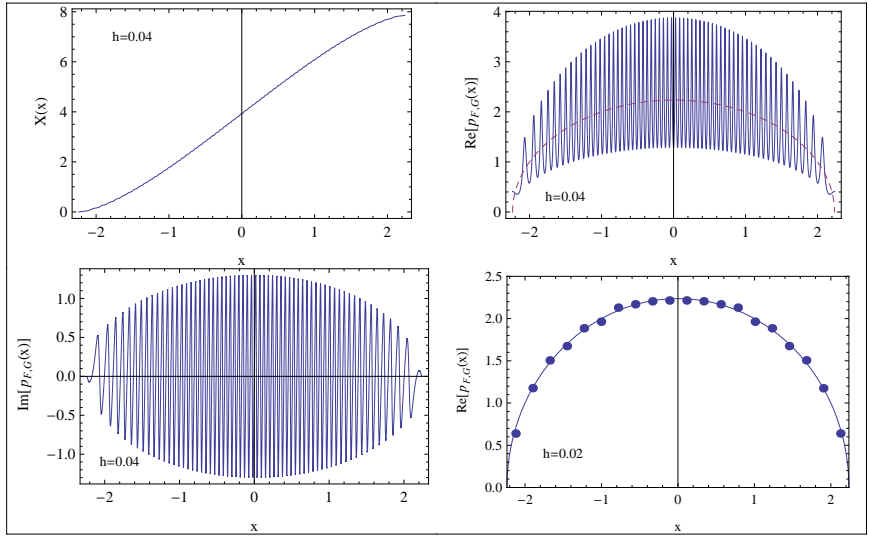}
\caption{The semiclassical limit. In the two upper boxes, the real parts of the quantum characteristic function (left box) and of the quantum momentum function (right box) for the energy $E = 2.5$, computed with a value of $\hbar = 0.04$ are reported. The dashed line in the right box is the classical momentum at the same energy. The imaginary part of the quantum momentum function is presented in the left lower box. In the lower right box is plotted the result of a coarse graining procedure on the real part of $p_{F,G}(x)$, computed with $\hbar =0.02$: the dots represents the mean values of this function, obtained averaging it in 20 small subintervals of the $x$-axis. The continuous line is the graph of the classical momentum. }
\end{figure}

In the two upper boxes, the real part of the quantum characteristic function and of the corresponding quantum momentum function for the energy $E= 2.5$, with a value of $\hbar = 0.04$ are reported. The imaginary part of the quantum momentum function is plotted in the left lower box. These figures should be compared with the corresponding ones for $n=2$ in Figs 1, 2 and 4, where $\hbar =1$.  As clearly seen from the figures, while decreasing $\hbar$, the real part of the quantum reduced action tends to become more and more close to the classical action $W_C(x)$, but maintaining a waving behavior around this latter, consequently acquiring an infinite number of ripples in the limit. Its derivative, i.e. the real part $Re[p_{F,G}(x)]$ of the quantum momentum, therefore presents an increasing number of oscillations of finite heights whose maximum amplitude tends to become constant while decreasing $\hbar$. The imaginary part of the quantum momentum function also presents an increasing number of fluctuations of finite amplitude around zero. The presence of a number of peaks and oscillations going to infinite suggests that these functions cannot directly tend in a strict mathematical sense to the corresponding classical quantities, but it is previously necessary to eliminate the peaks and oscillations by means of a coarse graining procedure. In the lower right box is plotted the result of such operation on the real part of $p_{F,G} (x)$: the dots represents the mean values of this function, in 20 small subintervals of the x-axis. The figure refers to the value $\hbar =0.02$. As seen from the figure, the dots are clearly distributed along the curve which represents the classical momentum $p_C(x)$.
This confirms that in order to obtain the classical quantities from the corresponding quantum ones, these latter have to be previously smoothed by eliminating the quantum fluctuations and are these smoothed quantities that tend to the classical ones for $\hbar\to 0$. As for the imaginary part of $p_{F,G} (x)$, it fluctuates around zero, and is furthermore proportional to $\hbar$ according to Eq. (14), so that its mean values tend to zero and the quantum momentum function for $\hbar \to 0$ in the c.a.r. becomes purely real, as should be. Also the quantum reduced action, in the c.a.r. becomes real, as can be seen from Eq. (15), by remembering that $X'(x)$ has no zeroes. The same conclusions are found for the hydrogen atom. The fluctuating behavior of the quantum reduced action and of the quantum momentum function in the limit $\hbar \to 0$ explains the difficulties of the WKB expansion.

\section{CONCLUSIONS}

The results presented in this paper prove that for one dimensional or reducible potentials the quantum Hamilton Jacobi equation is equivalent to the Schrodinger equation, offering an independent, alternative approach to quantum mechanics. Indeed, this equation can be postulated, as is usually done for the SE. The QHJE allows to identify and investigate the quantum characteristic function $W_G(x,E)$, that is the quantum analogue of the classical characteristic function $W_C(x,E)$, and the quantum momentum function $p_{F,G} (x,E)$ that is the analogue of the classical momentum. These quantities are complex in the classically allowed regions between the turning points, and imaginary outside. When $\hbar\to 0$, the quantum quantities in the c.a.r. become real and generate the corresponding classical ones.  This limit process can be investigated by means of the QHJE, clarifying the transition from the quantum to the classical mechanics. By means of this approach an exact WKB-like representation of the wave function is also found, which cannot be obtained from the SE. In this representation, the phase of the wave function is given by the real part of the quantum reduced action, while its amplitude is regulated by the derivative of this function. This shows that the reduced action is a fundamental quantity, both in quantum and classical mechanics. It is also interesting to note that the turning points of the classical mechanics have an important role in quantum mechanics, too. The method exposed here can be applied to other potentials, in particular to the study of the motion of a charged particle in the presence of a magnetic field, and to relativistic equations, problems which are presently under investigation, as well the modifications needed to study non-separable problems.

\end{document}